\documentclass{article}
\usepackage{spconf,amsmath,graphicx,hyperref,indentfirst,multirow,booktabs,makecell,enumitem,caption,subcaption,adjustbox}
\usepackage{hyperref}
\usepackage{etoolbox}
\usepackage{microtype}

\apptocmd{\thebibliography}{\setlength{\itemsep}{0pt}}{}{}

\title{FxSearcher: gradient-free text-driven audio transformation}
\name{Author(s) Name(s)\thanks{Thanks to XYZ agency for funding.}}
\name{Hojoon Ki \qquad Jongsuk Kim \qquad Minchan Kwon \qquad Junmo Kim}
\address{Korea Advanced Institute of Science and Technology, KAIST}

\begin{document}

\maketitle
\begin{abstract}
Achieving diverse and high-quality audio transformations from text prompts remains challenging, as existing methods are fundamentally constrained by their reliance on a limited set of differentiable audio effects. This paper proposes \textbf{FxSearcher}, a novel gradient-free framework that discovers the optimal configuration of audio effects (FX) to transform a source signal according to a text prompt. Our method employs Bayesian Optimization and CLAP-based score function to perform this search efficiently. Furthermore, a guiding prompt is introduced to prevent undesirable artifacts and enhance human preference. To objectively evaluate our method, we propose an AI-based evaluation framework. The results demonstrate that the highest scores achieved by our method on these metrics align closely with human preferences. Demos are available at \url{https://hojoonki.github.io/FxSearcher/}.
\end{abstract}

\section{Introduction}
\label{sec:intro}
Text-driven audio transformation aims to modify an existing audio signal based on a natural language description. 
One prominent approach involves using black-box neural networks (NNs), which offer the potential for a wide spectrum of transformations~\cite{li2023freevc, kim2024speak}. At the same time, this flexibility comes at the cost of being non-interpretable~\cite{steinmetz2021steerable} and risking corruption of the original signal~\cite{kong2020hifi}.
A contrasting approach leverages Audio Effects (FX) to transform audio, which is a class of signal processors that includes equalizers, reverbs, and distortions. This method offers deterministic and interpretable control of audio through mathematical transformations, mirroring the workflow of sound engineers~\cite{balasubramaniam2023word, steinmetz2022style}.

Recent research on FX control is categorized into two main approaches. The first is a gradient-based method, influenced by Differentiable Digital Signal Processing (DDSP)~\cite{engel2020ddsp, hayes2024review}. This allows FX to be integrated into deep learning frameworks, enabling end-to-end training via gradient descent. However, due to the differentiability constraint, these methods cannot fully reproduce the complex dynamics of commercial-grade FX~\cite{ramirez2021differentiable, kuznetsov2020differentiable}, potentially limiting expressive power~\cite{engel2020ddsp, hayes2024review, hayes2021neural}. For instance, while Text2FX~\cite{chu2025text2fx} effectively implements FX using gradient descent optimization, its reliance on differentiability confines it to a limited subset of FX like equalizers and reverbs, resulting in a lack of sonic diversity. The other approach~\cite{doh2025can} bypasses the gradient constraint by leveraging the rich knowledge of Large Language Models (LLMs) to generate FX parameters. Yet, these models operate solely on the text modality, disregarding the source audio entirely.

\begin{figure}[!t]
    \centering
    \includegraphics[width=\columnwidth]{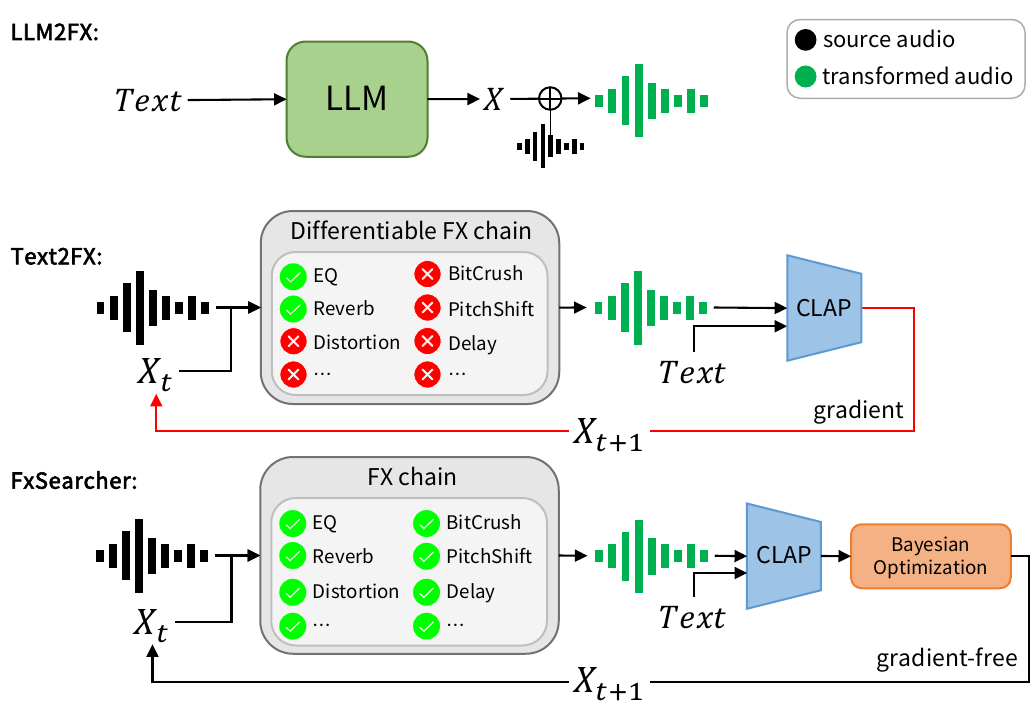}
    \caption{Comparison of operational mechanisms between other FX parameter generation models and our method, FxSearcher. }
    \label{fig:figure1}
\end{figure}
\begin{figure*}[!t]
    \centering
    \includegraphics[width=\textwidth]{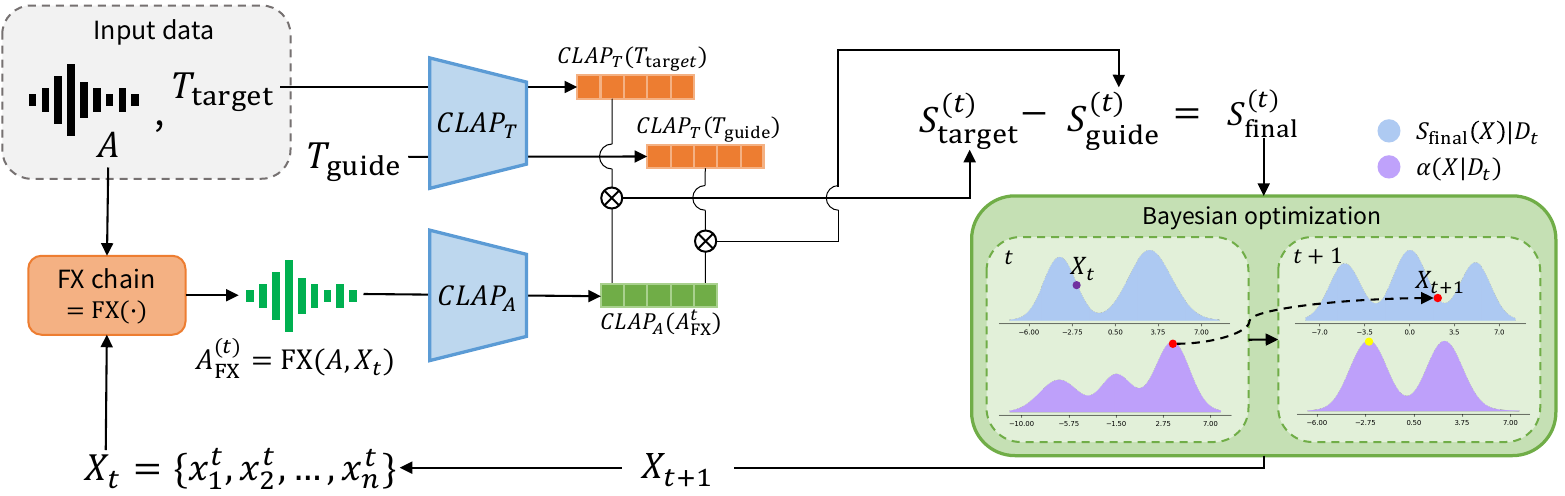}
    \caption{Overview of the proposed pipeline. The framework takes a source audio and a target prompt as input, generating a transformed audio and its corresponding optimal parameter set via an iterative optimization process.}
    \label{fig:figure2}
\end{figure*}

To address these issues, we propose ~\textbf{FxSearcher}, a novel framework that employs a gradient-free optimization approach. As depicted in Figure \ref{fig:figure1}, our framework enables the integration of any audio FX regardless of their differentiability. To achieve this, we employ Bayesian Optimization~\cite{jones1998efficient}, efficiently navigating the parameter space of audio FX. The optimization is guided by a score function based on the CLAP~\cite{elizalde2023clap}, which measures the semantic similarity between the transformed audio and a target text prompt. Since we find that optimizing for the target text prompt only often produces overly processed audio, we enhance the score function by introducing a guiding prompt strategy. Furthermore, we propose a set of AI-based objective metrics to validate the performance of our method. Combining these metrics with a Mean Opinion Score (MOS) test provides a more comprehensive and robust evaluation of the transformed audio. The main contributions of our paper are as follows:
\begin{enumerate}[nosep]
\item To the best of our knowledge, we propose the first gradient-free optimization framework for text-driven audio transformation.
\item We demonstrate that the guiding prompt strategy acts as an effective regularizer, preventing undesirable artifacts and leading to human-preferable results.
\item Our method achieves the highest scores across both our newly proposed AI-based objective metrics and standard human listening tests.
\end{enumerate}
\section{PROPOSED METHOD}
\label{sec:format}
\subsection{Overview}
The goal of our framework is to identify optimal audio effect parameters using natural language descriptions and audio. As illustrated in Figure \ref{fig:figure2}, FxSearcher operates in a closed-loop optimization driven by a Bayesian Optimization~\cite{jones1998efficient, frazier2018tutorial}. In each iteration, the algorithm proposes candidate parameters that are applied to a source audio by a predefined Audio FX Chain~\cite{SonicBloom2017EffectsChain}. The resulting audio is then evaluated by a score function utilizing CLAP~\cite{elizalde2023clap}. This function measures semantic alignment with a target prompt describing the desired attributes, and a guiding prompt is used to steer the optimization toward a more human-preferable sonic direction. The final score provides feedback to the Bayesian Optimization algorithm, guiding its subsequent search for a better solution. 

\subsection{Problem Definition}
Given a source audio $A$ and a target text prompt $T_{\text{target}}$, our primary objective is to find an optimal set of FX parameters, $X^*$, and the corresponding transformed audio, $A_\text{FX}^*$. 
Our pipeline iteratively optimizes the parameters of an audio FX chain. This sequential process can be formally represented as a single function, $\operatorname{FX}(\cdot)$. At iteration $t$, the chain takes the source audio $A$ and a set of parameters $X_t$ to generate the transformed audio $A_\text{FX}^{(t)}$ as
\begin{equation}
    A_\text{FX}^{(t)} = \operatorname{FX}(A, X_t).
\end{equation}
\subsection{Score Function}
Our score function is designed to guide the optimization toward results that are not only textually relevant but also human-preferable. For this, we compute a holistic score from two complementary prompts. The primary \textbf{target prompt}, $T_{\text{target}}$, provides the literal description of the desired sound.
A \textbf{guiding prompt}, $T_{\text{guide}}$, which is a description of common undesirable artifacts (\textit{A harsh, distorted, muddy, unclear, oversaturated, unpleasant sound.}), is used to refine the search and steer the result toward a more human-preferable quality.
Based on these prompts, we calculate a \textbf{target score} ($S_{\text{target}}$) and a \textbf{guiding score} ($S_{\text{guide}}$) using CLAP as follow:
\begin{align}
S_{\text{target}}^{(t)} &= \text{sim}(CLAP_{A}(A_\text{FX}^{(t)}), CLAP_{T}(T_{\text{target}})), \\
S_{\text{guide}}^{(t)} &= \text{sim}(CLAP_{A}(A_\text{FX}^{(t)}), CLAP_{T}(T_{\text{guide}})),
\end{align}
where $\text{sim}(\cdot, \cdot)$ denotes the cosine similarity, and $CLAP_{A}$ and $CLAP_{T}$ represent the audio and text CLAP encoders, respectively. 
To simultaneously increase similarity to the target description while decreasing similarity to undesirable characteristics, we aim to maximize the difference between the target and guiding scores. Therefore, the final objective score $S_\text{final}^{(t)}$ is formulated as:
\begin{equation}
S_\text{final}^{(t)} = S_\text{target}^{(t)} - S_\text{guide}^{(t)}.
\end{equation}

\subsection{Optimization}
To find the optimal parameters $X^*$, we employ Bayesian Optimization~\cite{jones1998efficient}, which is well-suited for optimizing black-box objective functions~\cite{snoek2012practical}. It works by building a probabilistic surrogate model of the objective function using a Gaussian Process ($\mathcal{GP}$). After $t$ iterations, with the observed data $D_t = \{(X_1, S_\text{final}^{(1)}), ..., (X_t, S_\text{final}^{(t)})\}$, the model allows us to predict the score for any candidate parameter set $X$ in the entire search space:
\begin{equation}
S_\text{final}(X) | D_t \sim \mathcal{GP}(\mu_t(X), k_t(X, X')).
\end{equation}
The mean function $\mu_t(X)$ represents the best prediction for the score at point $X$, and a covariance function $k_t(X, X')$ models the uncertainty by describing the correlation between the scores at two arbitrary points, $X$ and $X'$.
 
The surrogate model informs the strategic search by providing a score prediction via its mean function and a measure of confidence derived from its covariance function. An acquisition function, $\alpha(X)$, leverages this information to balance two objectives: exploitation, the sampling of promising regions with high predicted scores, and exploration, the sampling of regions with high uncertainty. Maximizing the acquisition function determines the most promising point to evaluate next. This can be defined as
\begin{equation}
X_{t+1} = \underset{X}{\arg\max} \, \alpha(X | D_t).
\end{equation}
With the new FX parameter set $X_{t+1}$, the optimization process proceeds to the next iteration. The pipeline repeats the process, iteratively converging on the optimal parameter set $X^*$ and the final transformed audio $A_\text{FX}^*$.
\section{EXPERIMENTS}
\label{sec:experiment}
\begin{figure}[!t]
    \centering
    \includegraphics[width=\columnwidth]{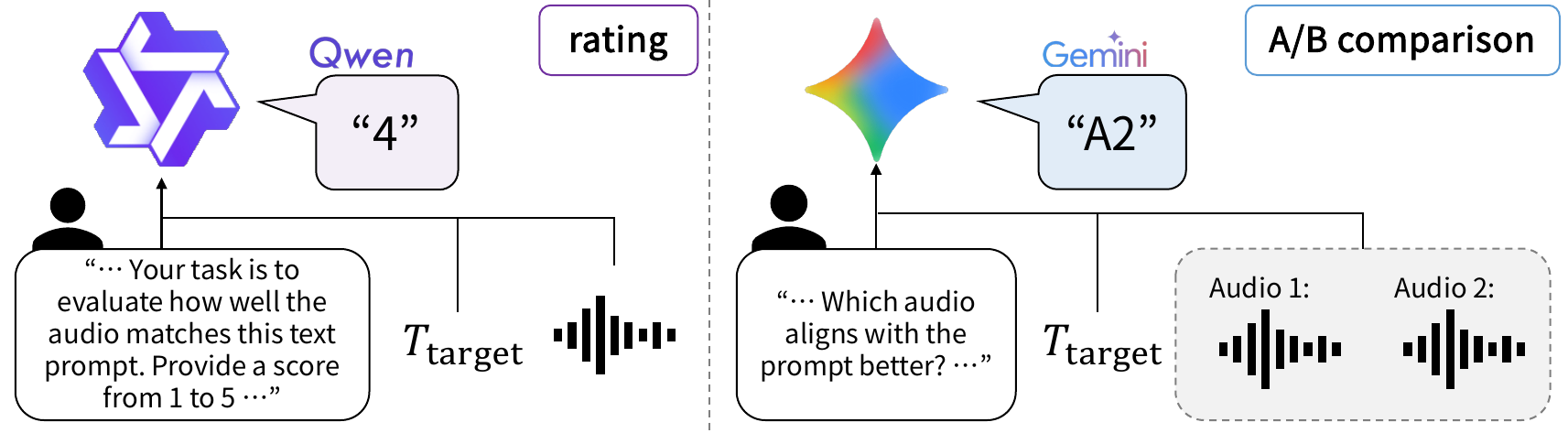}
    \caption{Illustration of the two AI-based evaluation methods. \textbf{Left}: Qwen providing a 1-5 absolute rating for a single audio based on a prompt. \textbf{Right}: Gemini performing a pairwise A/B comparison to choose the better of two audio.}
    \label{fig:figure3}
\end{figure}
\subsection{Experimental Details}
\noindent\textbf{FX Chain.} The FX chain is configured to apply a total of six effectors sequentially, using Spotify's Pedalboard ~\cite{sobot_peter_2023_7817838} with an order determined by typical signal flow in sound engineering~\cite{SonicBloom2017EffectsChain}: Equalizer $\rightarrow$ Distortion $\rightarrow$ BitCrush $\rightarrow$ PitchShift $\rightarrow$ Delay $\rightarrow$ Reverb. The Equalizer first shapes the tone, followed by Distortion for adding texture. BitCrush and PitchShift then introduce more dramatic changes, while Delay and Reverb finalize the sound by imparting a sense of space. The audio FX chain is controlled by 26 parameters, which are divided into two main types. The first group consists of 22 parameters dedicated to configuring the effect settings: 15 for the Equalizer, 3 for Reverb, and 1 for each of the four other effects. The second group contains the 4 remaining parameters, which control the activation (on/off) of those four optional effects.


\begin{table}[!t]
    \centering
    \small
    \setlength{\tabcolsep}{2pt}
    \begin{adjustbox}{max width=\linewidth}
    \begin{tabular}{llc|cccc} 
        \toprule
        \multirow{2}{*}{\textbf{Domain}} & \multirow{2}{*}{\textbf{Method}} & \multirow{2}{*}{\textbf{Time (s)}} & \multicolumn{4}{c}{\textbf{Evaluation Metric}} \\
        \cmidrule(l){4-7} 
         & & & \textbf{CLAP} & \textbf{MOS} & \textbf{QWEN} & \textbf{Gemini-WR} \\
        \midrule
        \multirow{3}{*}{Speech} & LLM2FX & 71.7 & 0.232 & 1.77 & 2.32 & 38.2 \\
         & Text2FX & 197.4 & \textbf{0.527} & 2.28 & 2.38 & \textbf{51.3} \\
         & \textbf{FxSearcher} & \ 71.9 & 0.447 & \textbf{3.48} & \textbf{2.73} & \textbf{61.8} / 48.7 \\
        \midrule
        \multirow{3}{*}{Instrumental} & LLM2FX & \ 71.9 & 0.341 & 2.70 & 3.14 & 28.4 \\
         & Text2FX & 165.5 & \textbf{0.561} & 3.19 & 3.03 & 33.8 \\
         & \textbf{FxSearcher} & \ 71.9 & 0.464 & \textbf{3.46} & \textbf{3.18} & \textbf{71.6 / 66.2} \\
        \bottomrule
    \end{tabular}
    \end{adjustbox}
    \caption{Comparison of different methods. Time is measured in seconds per sample. Gemini-WR of FxSearcher represents win rates against (LLM2FX) / (Text2FX), respectively.}
    \label{tab:table1}
\end{table}

\noindent\textbf{Dataset.} We construct a dataset consisting of source audio and target prompts. Audio samples for the speech domain and instrumental domain are drawn from LibriSpeech~\cite{panayotov2015librispeech} and a public dataset\footnote{\url{https://www.kaggle.com/datasets/abdulvahap/music-instrunment-sounds-for-classification}}, respectively. We prepare 150 text prompts for the evaluation. 120 prompts are automatically generated using GPT-5~\footnote{\url{https://openai.com/index/introducing-gpt-5/}}-60 for the speech domain and 60 for the instrumental domain, while the remaining 30 are designed by the researchers.

\noindent\textbf{Configuration.}
For our experiments, the maximum number of search iterations is set to 100, and the early stopping patience is set to 30. For the comparison, Text2FX~\cite{chu2025text2fx} and LLM2FX~\cite{doh2025can} are reproduced in our environment. Also, we utilize a pretrained CLAP model~\footnote{\url{https://huggingface.co/laion/clap-htsat-unfused}} for optimization and evaluation. All experiments are conducted on a single NVIDIA RTX 3090 GPU.
\subsection{Evaluation Metrics}
To evaluate generated audio, we employ a combination of objective and subjective metrics.

\noindent\textbf{CLAP Score.} We use the CLAP~\cite{elizalde2023clap} score to evaluate the alignment between the text prompt and the transformed audio.

\noindent\textbf{Mean Opinion Score (MOS).} For subjective human evaluation, we conduct a Mean Opinion Score (MOS) test. A total of 16 participants rated the \textit{alignment between text and audio} on a scale from 1 to 5.

\noindent\textbf{AI-based Evaluation.} As illustrated in Figure~\ref{fig:figure3}, we introduce two distinct AI-based methods for a comprehensive assessment. For a direct, quantitative rating of text-audio alignment, we use the QWEN score (Qwen2.5-omni-7B~\cite{xu2025qwen2}) on a 1-5 scale. 
To conduct a more nuanced, relative evaluation, we opt for a pairwise A/B comparison. This task requires the model to be capable of processing two audio inputs simultaneously for a direct comparison. Therefore, we leverage the Gemini 2.5 Flash API~\cite{comanici2025gemini}, as it provides the reliable performance needed for this capability and is an affordable solution.

\noindent\textbf{Objective Audio Quality Metrics.} We measure Fréchet Audio Distance (FAD)~\cite{kilgour2018fr} for distributional similarity and Loudness~\cite{series2011algorithms} and its standard deviation for stability. For speech, we add Word Error Rate (WER, via Whisper-large-v3~\cite{radford2023robust}) and Perceptual Evaluation of Speech Quality (PESQ)~\cite{recommendation2001perceptual} for intelligibility.

\subsection{Main Result}
\begin{table}[!t]
    \centering
    \small
    \setlength{\tabcolsep}{1.5pt}
    \begin{tabular}{l ccccccc}
        \toprule
        \multicolumn{1}{c}{\multirow{2}{*}{Method}} & \multicolumn{2}{c}{Speech} & \multicolumn{4}{c}{Overall} \\
        \cmidrule(r){2-3} \cmidrule(l){4-7}
         & WER $\downarrow$ & PESQ$\uparrow$ & FAD$\downarrow$ & CLAP$\uparrow$ & MOS$\uparrow$ & Gemini-WR$\uparrow$\\ 
        \midrule
        \textbf{FxSearcher} & \textbf{37.5} & \textbf{1.09} & \textbf{11.23} & 0.456 & \textbf{3.47} & \textbf{51.3}\\
        w/o $T_{\text{guide}}$ & 53.0 & 1.06 & 14.19 & \textbf{0.482} & 2.99 & 48.7\\ 
        \bottomrule
    \end{tabular}
    \caption{Ablation study of guiding prompt on FxSearcher performance across speech and overall quality metrics.}
    \label{tab:table2}
\end{table}
\begin{figure}[!t]
    \centering 

    \begin{minipage}[c]{0.48\columnwidth}
        \centering
        \includegraphics[width=\linewidth]{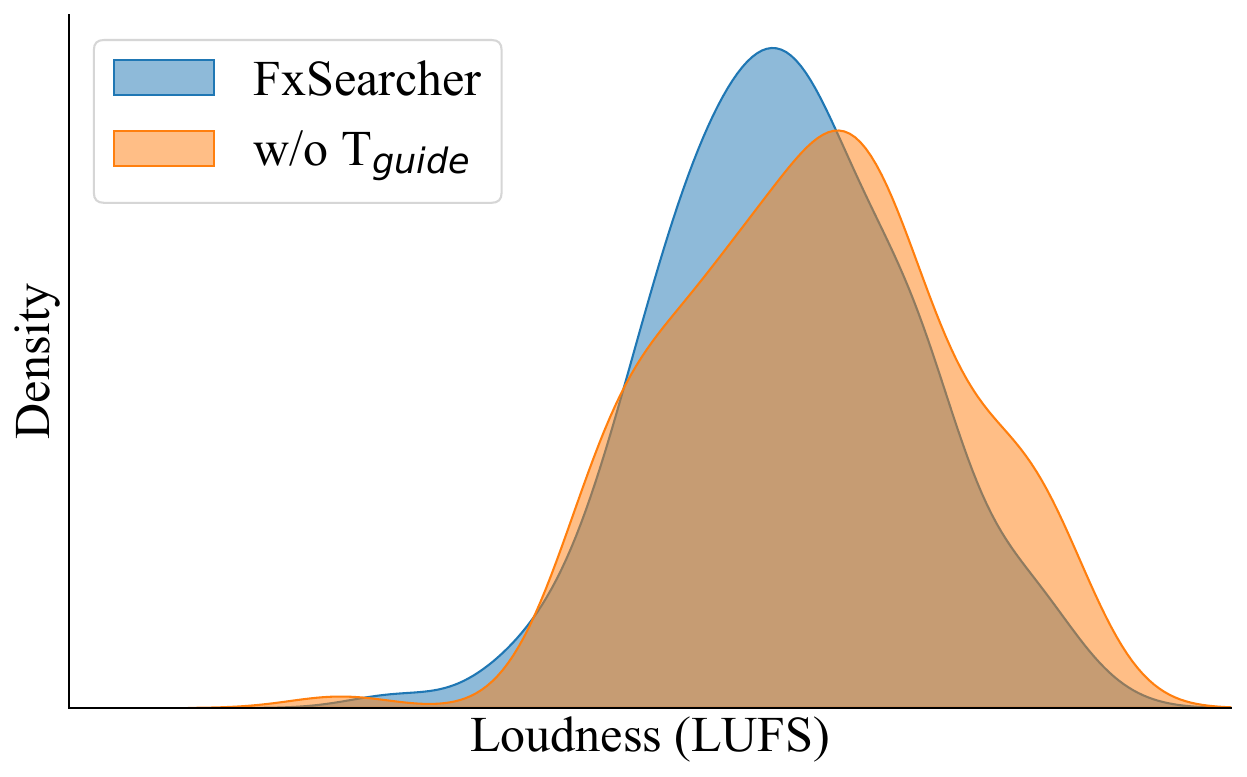}
        \label{fig:my_diagram}
    \end{minipage}
    \begin{minipage}[c]{0.48\columnwidth}
        \centering
        \footnotesize
        \setlength{\tabcolsep}{2pt}
        \begin{tabular}{lcc}
        
            \toprule
            \textbf{Method} & \textbf{Loudness} & \textbf{Std} \\
            \midrule
            FxSearcher & -20.70 & 8.57 \\
            - w/o $T_\text{guide}$ & -18.88 & 9.52 \\
            \bottomrule
        \end{tabular}
        \label{tab:sub_loudness}
    \end{minipage}
    \vspace{-10pt}
    \caption{Effect of the guiding prompt on loudness distribution. The plot (left) shows the density of loudness values (in LUFS), while the table (right) summarizes their mean and standard deviation.}
    \label{fig:figure4}
\end{figure}
The performance comparison between FxSearcher and the baseline is summarized in Table~\ref{tab:table1}. 
In terms of computational efficiency, FxSearcher shows a significant advantage over Text2FX and operates at a comparable speed to LLM2FX. 
Regarding the objective CLAP score, Text2FX achieves the highest, while LLM2FX yields the lowest. 
The poor performance of LLM2FX is attributed to its operational mechanism, which does not consider the input audio. 
The strong performance of Text2FX is anticipated, as its gradient-based method is explicitly designed to maximize this single metric. 
Black-box approaches like ours, in contrast, explore the parameter space more broadly without such direct guidance~\cite{nesterov2013introductory, banker2025gradient}.

In the human evaluation (MOS), FxSearcher achieves a significantly higher score, indicating that its results are more perceptually aligned with user expectations. We further validate this finding through our AI evaluations. FxSearcher achieves the highest scores on both AI-based metrics: the QWEN score and the win rate from Gemini 2.5 Flash. We observe that the correlation between the CLAP score and human preference diminishes at higher score ranges. This combined evidence suggests that while the CLAP score is a useful guide, it serves as an incomplete proxy for perceptual quality. Therefore, by employing the suite of AI-based metrics that we have introduced, it is possible to achieve results that more accurately reflect human perception.

\subsection{Ablation Study}
\noindent\textbf{Guiding Prompt.} Table~\ref{tab:table2} shows the effectiveness of the guiding prompt. Using the guiding prompt significantly improves speech clarity and quality (lower WER, higher PESQ), and prevents excessive deviation from the original audio (lower FAD). 
The acoustic stability is also enhanced, as detailed in Figure~\ref{fig:figure4}. 
The guiding prompt leads to a smoother and more predictable distribution, as evidenced by its lower mean and standard deviation.
The raw target score ($S_{\text{target}}$) is higher without the guiding prompt, as the optimization is focused on a single objective. 
While an AI judge shows a slight preference for the guided result, the effect on human listeners is decisive. 
The MOS evaluation confirms a significant preference for audio generated with the guide, validating its direct impact on perceptual quality. 
\begin{table}[!t]
\centering
\small
\setlength{\tabcolsep}{2pt}
\begin{tabular}{lcccc}
\toprule
\multicolumn{1}{c}{\multirow{2}{*}{\textbf{FX Chain}}} & \multicolumn{2}{c}{\textbf{CLAP}} & \multicolumn{2}{c}{\textbf{QWEN}} \\
\cmidrule(lr){2-3} \cmidrule(lr){4-5}
& Speech & Instrumental & Speech & Instrumental \\
\midrule
Equalizer $\rightarrow$ Reverb & 0.389 & 0.428 & 2.32 & 3.11 \\
+ Distortion & 0.397 & 0.439 & 2.31 & 3.14 \\
+ BitCrush & 0.409 & 0.437 & 2.45 & 3.16 \\
+ PitchShift & 0.445 & 0.457 & 2.62 & 3.15 \\
+ Delay(Ours) & \textbf{0.447} & \textbf{0.464} & \textbf{2.73} & \textbf{3.18} \\
\bottomrule
\end{tabular}
 \caption{Results of the ablation study on the FX chain. The final configuration(+Delay) represents our full FxSearcher framework.}
    \label{tab:table3}
\end{table}

\noindent\textbf{FX Chain.} Table~\ref{tab:table3} investigates how the diversity of the available FX pool impacts performance. We start with a default chain (Equalizer→Reverb) and progressively expand it by sequentially adding distinct effects. All effects used in this chain are standard, indifferentiable versions. 
The results show a clear, monotonic improvement as the FX chain grows. With each new effect introduced, both the CLAP and QWEN scores consistently increase. This performance gain is attributed to two main factors. The first is the isolated contribution of each component, which is especially pronounced upon the inclusion of PitchShift. The second factor is the richness of the expanded sonic palette, which provides more degrees of freedom, enabling FxSearcher to discover more effective and nuanced parameter combinations to satisfy the target prompt.
\section{CONCLUSIONS}
\label{sec:conclusions}

In this study, we introduce FxSearcher, a novel gradient-free framework that controls FX to transform audio from text prompts. The strong preference for our method in both human and AI evaluations stems from its core advantage: compatibility with any FX plugin, which unlocks a far greater sonic diversity. We also demonstrated that the guiding prompt strategy improves audio quality and stability. We expect FxSearcher to become an intuitive tool that can democratize complex audio editing for experts and novices alike.

{\small
\bibliographystyle{IEEEbib}
\bibliography{refs}
}
\end{document}